# An Efficient Attention Mechanism for Sequential Recommendation Tasks: HydraRec

Uzma Mushtaque[1]

[1]Rensselaer Polytechnic Institute, Troy, USA
`mushtu@rpi.edu`

**Abstract.** Transformer based models are increasingly being used in various domains including recommender systems (RS). Pretrained transformer models such as BERT have shown good performance at language modelling. With the greater ability to model sequential tasks, variants of Encoder-only models (like BERT4Rec, SASRec etc.) have found success in sequential RS problems. Computing dot-product attention in traditional transformer models has quadratic complexity in sequence length. This is a bigger problem with RS because unlike language models, new items are added to the catalogue every day. User buying history is a dynamic sequence which depends on multiple factors. Recently, various linear attention models have tried to solve this problem by making the model linear in sequence length (token dimensions). Hydra attention is one such linear complexity model proposed for vision transformers which reduces the complexity of attention for both the number of tokens as well as model embedding dimensions. Building on the idea of Hydra attention, we introduce an efficient Transformer based Sequential RS (HydraRec) which significantly improves theoretical complexity of computing attention for longer sequences and bigger datasets while preserving the temporal context. Extensive experiments are conducted to evaluate other linear transformer-based RS models and compared with HydraRec across various evaluation metrics. HydraRec outperforms other linear attention-based models as well as dot-product based attention models when used with causal masking for sequential recommendation next item prediction tasks. For bi-directional models its performance is comparable to the BERT4Rec model with an improvement in running time.

**Keywords:** Information Retrieval, Sequential Recommendations, Linear attention, BERT4Rec, HydraRec, Transformer Models

## 1    Introduction

For most online services, recommendations allow an effective strategy for exposing users to relevant items. In most scenarios this relevance is computed using user's past interactions with various items. User preferences evolve over time. The sequence in which items are bought reveals a buying pattern for every user. For example, a user who bought a computer might need to buy accessories for the computer. Identifying temporal semantics of user interactions in a sequence and using that to make future predictions is termed as long-term sequential recommendation (LSR). Various models have been used to capture sequential dynamics of user behavior [1], [2]. Most of these models predict the next item given a sequence of historical user interactions. Recurrent neural networks (RNNs) and their variants have been widely used for sequential recommendations. These models encode user-item interactions in a unidirectional manner. This encoding/hidden representation is then used to predict the next item or items [3].

Attention [4] based approaches using bi-directional networks like BERT [5] exhibit superior performance by learning encodings bi-directionally. Originally these models were introduced for natural language processing (NLP) related tasks and time series problems. However, given the similarity of LSR with NLP problems, variants of NLP models are used in recommender system applications [2]. These models are modified to specifically solve recommender system problems of next item prediction and next basket prediction. Despite the similarity between NLP tasks and LSR, there are some fundamental differences. The general idea of these models is to consider items as words in a sequence like NLP models and represent each item as a token. The basic building block of these models is the dot-product attention mechanism that calculates the attention matrix for different items based on their relevance in the sequence. One limitation of these models is the complexity of the dot product operation for sequential recommendation tasks which is quadratic in the number of tokens(items). This becomes an issue when the sequence length ($N$) is much greater than the item embedding size ($d$). Several approaches have been proposed to deal with this issue such as Fixed Patterns (FP), Combination of Patterns (CP), Learnable Patterns (LP), Low-Rank Methods, Kernels etc. [6]. Various transformers with linear complexity have also been introduced [7][8]. When dealing with LSRs these methods exhibit low accuracy [9]. The LinRec model [10] addresses these shortcomings by introducing L2-normalized linear attention for long-term sequential recommendations. This model changes the order of operations to calculate attention. Flash attention model demonstrated superiority over these models even using a few thousand tokens and achieving reasonable performance [12].

There is a trade-off between efficiency and accuracy which is not specifically studied in the RS domain. Unlike NLP, LSR poses an additional modelling challenge because unlike language models, new items (movies, products etc.) are introduced almost every day which leads to additional tokens (item representations) being added to the model. There is a need for more efficient and accurate models that are linear in tokens as well as embedding (model) dimensions. Most of the 'linear' attention models trade computation across tokens for computation across embedding dimensions[11]. To the best of our knowledge, most of the linear models focus on model complexity with respect to either sequence length($N$) or model dimensions($d$). To address these research gaps, we introduce a novel transformer-based model called HydraRec that is linear in both sequence length and model dimensions. HydraRec significantly improves computational efficiency for the LSR problem. This model is built on the idea of Hydra Attention [11], which introduces an attention framework that is linear in model dimensions as well as sequence length for vision tasks. It is known that Vision Transformer (ViT) space requires more efficient models, and hence many variants of attention exist [14][15]. ViT problems require modeling of special temporal features efficiently which is remarkably similar to the sequential recommendation problem. In the current work we take inspiration from vision transformers and introduce HydraRec which is an encoder-based transformer model built on the principle of hydra attention model [11]. This model considers linear attention and the impact of increasing number of heads equal to the number of features such that the order complexity becomes linear in tokens and embedding dimensions. Like the original model we explore various kernels for increasing accuracy in the cloze task [13]. *An important aspect of HydraRec is its ability to create a global feature vector across the entire given item sequence and then filtering the importance of this global feature for each output. This makes it different from other linear models.*

To preserve temporal context of longer sequences bi-directional models are found to be effective [13], therefore we use BERT4Rec architecture to strike a balance between accuracy and efficiency. In general, training a bi-directional model on item sequences causes information leakage because of its ability to 'see' the items of the future. BERT4Rec model solves this issue by training a model for cloze task [13]. BERT4Rec, calculates the context for each item bidirectionally. There are some unidirectional models like SASRec [2] that achieve good performance as well. In this work we consider both bidirectional and unidirectional context evaluation of attention and train the model for the cloze task (masked item prediction in a sequence).

The major contributions of this work are:

1) We developed a novel multi-head linear attention model for sequential recommendations called HydraRec. In our knowledge, this is one of the early works that uses the concept of vision transformers in sequential recommender systems.

2) HydraRec reduces the complexity of the Linear model to be Linear in token dimensions as well as embedding dimensions (for cases when the number of heads is equal to the model dimensions). This is an important concern for LSR based problems.

3) The proposed HydraRec mechanism can be applied to all Transformer-based models. In this study we incorporate it with BERT4Rec transformer-based architecture only. This is done to understand the impact of Hydra attention on a sequential recommendation task.

4) HydraRec is evaluated both under a unidirectional and bidirectional context calculation of attention.

5) Extensive experiments are conducted to study the trade-off between accuracy and efficiency of HydraRec with other Linear attention-based transformers RS models and the baseline model (BERT4Rec).

The rest of the paper is organized as follows: Section 2 describes the background of attention bases models along with their application in sequential recommendations and a literature review of existing work in the area. Section 3 describes the methodology including problem statement, model architecture and the core HydraRec model. This section describes how the model achieves linear complexity. Section 4 covers model training details including data description, experiments, results and discussion. The final section is section 5 which consists of conclusion and future work.

## 2      Background

In this section we discuss the dot-product attention mechanism widely used in Transformer based models [4]. Next, we describe the sequential recommendation problem and conduct a literature review of existing work.

### 2.1 Dot Product Attention

The most significant part of a Transformer model is the scaled dot-product attention [4], which captures the context of a given token w.r.t other tokens in the sequence. If the length of a sequence is denoted by $N$ and $d$ is the dimensionality of each token. The scaled dot-product attention ($A$) is given by:

$$A(Q, K, V) = softmax\left(\frac{QK^T}{\sqrt{d}}\right)V \qquad (1)$$

Here $X \in R^{N \times d}$ is the input matrix. Weight matrices $W_Q, W_K, W_V$ are learnt from the training process and $Q, K, V \in R^{N \times d}$ are query, key and value matrices. The softmax function is applied row-wise to the fraction $\frac{QK^T}{\sqrt{d}}$. $A$ is referred to as the attention matrix. The dimensionality of $A$ is fixed and the attention information can be transferred to the sequence easily. The disadvantage of the $QK^T$ calculation is the complexity associated with it which is quadratic in sequence length (number of tokens). This is an issue for problems involving LSR given the memory and time complexity associated.

### 2.2 Sequential Recommendations

Sequential Recommendation Systems (SRS) are the sequence of interactions a user makes in a chronological order over a specified period. SRS can be short or long. In general, if the item embedding is denoted by $d$ and the number of historical interactions is $N$, then the ratio $N/d$ greater than 1.5 is considered a long sequence [10]. In this study we focus on LSR, and the computational cost associated with the models that consider SRS. We do not restrict our experimentation to the ratio restriction of 1.5 because the goal of this study is to establish the effectiveness of HydraRec for all types of SRS tasks.

## 2.3 Related Work

SRS have been modelled by capturing user historical interactions via Markov chains (MCs) [16]. Later many hybrid approaches involving MCs were also introduced to solve the SRS problem [17] [18]. A more recent trend is to use sequential deep learning models like Recurrent Neural Networks (RNNs) and their variants for SRS. User interaction sequences can be used to model user behavior by using models like Gated Recurrent Units (GRU) and Long Short-Term Memory (LSTM) [19]. The general idea of these models is to efficiently combine past observations and create vector representations. Some of the recent works in this category are [20], [21], [22], [23]. In addition to RNNs, there are other deep learning models like Convolution Neural Networks (CNNs) and MLP networks that have been used to solve the problem of SRS [24], [25].

Transformer based models that use attention mechanism are found to be extremely effective in modelling sequential data [4][26]. Some methods have used attention with other models to model session-based recommendations [1]. Deviating from the complete transformer architecture are encoder [5] only and decoder only models [27].These models have been used for modeling SRS effectively. Most transformer models in NLP applications calculate dot product self-attention via causal masking. Many such models have found application in recommender system literature [2] [28]. Bi-directional frameworks like BERT[5] calculate the dot product bi-directionally taking context from both ends into account. One BERT based model used for SRS is BERT4Rec [13] which is an encoder only model and uses a multi-head self-attention mechanism. A significant limitation of these models when applied to long SRS is the computational and memory complexity. The dot product operation is quadratic in the number of tokens (vocabulary size in NLP models and number of items in RS models). This is a bigger problem for RS due to the ever-growing catalogue size. Linear attention models have tried to solve this problem via a decomposable kernel [7]. LinRec [10] applies linear attention to SRS problem and achieves a complexity that is linear in the number of tokens but quadratic in embedding dimensions.

Most retail and/or subscription platforms have items added to their catalogue every day and therefore user interaction sequences grow over time. For extremely large sequences it is required that the models are extremely efficient without losing semantics behind sequential interactions. Temporal dependencies are also modeled by vision transformers [29] and the goal is to maintain efficiency. One such vision transformer model is Hydra Attention model [11] that takes the linear attention calculation a step further by maximizing the number of attention heads resulting a model that is linear in token and embedding dimensions. The paper also introduces various kernels and discusses the effectiveness of the choice of various kernels in calculating linear attention. In this work, we build both unidirectional and a bi-directional encoder-based model that works on the principle of Hydra attention called HydraRec. Hydra attention is based on the idea of increasing the number of heads in multi-headed attention and its impact on the overall performance of the model. We build HydraRec on linear attention (like LinRec) and experiment with other kernels as well as number of heads. The original paper suggests replacing self-attention with different strategies. HydraRec only replaces SoftMax self-attention with Hydra attention. Additionally, our goal is to compare the performance of linear mechanisms when used for the SRS problem, therefore we keep the base model as BERT4Rec and experiment with various linear attention mechanisms.

## 3      Methodology

In this section we state the LSR problem mathematically, describe the HydraRec model, its architecture and discuss the training process.

### 3.1 Problem Statement

For any recommendation problem, the goal is to match a given set of users $U = \{u_1, u_2 \dots u_{|U|}\}$ to a given set of items $V = \{v_1, v_2 \dots v_{|V|}\}$. The SRS problem also involves user interaction history in chronological order. Let the interaction sequence for any user $u \in U$ given by $Seq_u = [v_1^u, \dots, v_t^u, \dots, v_{n_u}^u]$. Here $n_u$ is the interaction sequence length. More formally, the SRS problem is to predict the item that the user will interact with at time step $n_u + 1$. To solve this problem, we model the following probability for all possible items $v \in V$: $p(v_{n_u+1}^u = v | Seq_u)$.

### 3.2 Model Architecture

In the current work, we build the HydraRec model using Bert4Rec architecture only. The attention computation can be extended to any other transformer-based model. HydraRec model comprises of $L$ bi-directional transformer layers capable of sharing information across all positions (for item sequences) from the previous layer along with the transformer layer. We experiment with unidirectional model too by modifying the attention masking strategy by effectively creating a causal (left to right) attention mask.

### 3.2.1 Transformer Layer

Like many other transformer-based models, HydraRec computes the hidden representation $h_i^l$ for each token $i$ given sequence length $N$. Next, these hidden representations are stacked together into a matrix $H^l \in R^{N \times d}$. The attention function from equation (1) is computed on all positions simultaneously. These hidden representations are an input to the transformer layer which consists of two sub-layers: Multi-Head Self-Attention (MHSA) layer and Position-Wise Feed Forward (PFF) Network Layer. MHSA is an integral part of transformer architecture where each head ($h$) creates its own attention matrix. This can be represented as:

$MH(H^l) = [h_1; h_2; \dots; h_{head}]W^o$

$where \ h_i = Attention(H^l W_Q^i, H^l W_K^i, H^l W_V^i)$

Here all the weight matrices are learnable parameters. The hidden representations of each layer as well as item embeddings are built exactly as specified in [13]. The Hydra attention mechanism is built on the idea of increasing the number of heads in Linear attention mechanisms [11] [7]. HydraRec model uses Hydra attention for the LSR problem. In equation (1) dot product attention is computed. This is an $O(N^2 d)$ operation and scales poorly as the sequence length increases. Linear attention mechanisms generalize the softmax operation and use a decomposable kernel. A more general representation of attention is:

$$A(Q, K, V) = sim(Q, K)V \tag{2}$$

Here sim(.) is a similarity function. Any decomposable kernel $\varphi(.)$ can be used to express this similarity i.e. $sim(a, b) = \varphi(a)\varphi(b)^T$, then after using associativity equation (2) can be written as:

$$A(Q, K, V, \varphi) = \varphi(Q)(\varphi(K)^T V) \tag{3}$$

Here the operation $\varphi(K)^T V$ is computed first leading to a complexity that is linear in sequence length but quadratic in model dimension $O(Nd^2)$.

### 3.2.2 The HydraRec Model

Vision transformers have used different number of heads to better scale image related problems [11]. In the original attention model [4], each head $H$ has its own subset of features $d/H$, therefore equation (1) can be re-written as:

$$A(Q_h, K_h, V_h) = softmax\left(\frac{Q_h K_h^T}{\sqrt{d}}\right)V_h \quad \forall \in \{1, \dots, H\} \tag{4}$$

However, this does not impact the overall complexity of the model with dot-product attention. On the other hand, with Linear attention adding the number of heads decreases the number of operations because the complexity becomes $O\left(HN\left(\frac{d}{H}\right)^2\right) = O(\frac{Nd^2}{H})$ and equation (3) becomes:

$$A(Q_h, K_h, V_h, \varphi) = \varphi(Q_h)(\varphi(K_h)^T V_h) \quad \forall h \in \{1, \dots, H\} \tag{5}$$

In (5), $Q_h, K_h, V_h$ are column vectors with dimensionality $R^{N \times 1}$. Vectorizing the operation across heads gives hydra attention:

$$Hydra(Q, K, V, \varphi) = \varphi(Q)\sum_{i=0}^{N}\varphi(K)^T V^T \tag{6}$$

The multiplication $\varphi(K)^T V^T$ is elementwise multiplication. The final summation multiplication with $\varphi(Q)$ is also elementwise. Hydra attention is different from the original scaled dot-product attention because it creates a more generic feature vector for the entire input sequence with the $\sum_{i=0}^{N}\varphi(K)^T V^T$ operation. Multiplication with $\varphi(Q)$ is equivalent to filtering the relevance of each token (item) in the sequence. This is a more generic approach towards modeling contextual information within a sequence. The complexity of this model is $O\left(Nd\left(\frac{d}{H}\right)\right)$. Here the idea is that increasing the number of heads decreases the number of operations. If $H = d$, then the model complexity becomes $O(Nd)$. This is true for both time and space complexity. We incorporate this attention mechanism in one of the well-known encoder based architectures specifically used for the LSR problem – BERT4Rec [13] and call this model HydraRec. It is worth noting that this attention framework can be utilized for any transformer-based RS architecture (Figure 1). For the overall model framework no changes are made to the architecture specified in [13].

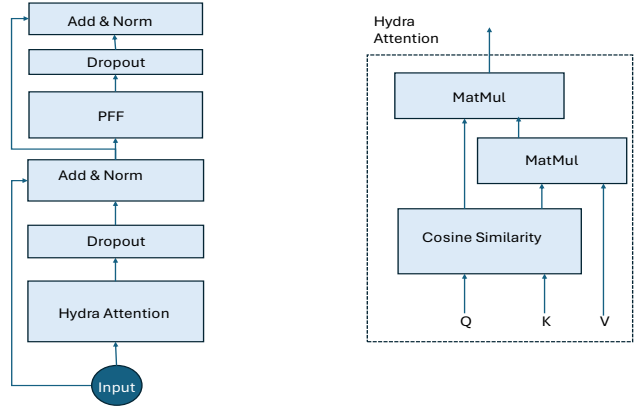

**Fig. 1** **Architecture of HydraRec within the Transformer Layer for any Attention based Model**

## 4       Model Training

The overall model architecture uses PFF which is a concatenation of Feed forward Network (FFN) applied to the output of the attention layer described above. The FFN layer comprises of two affine transformations with a GELU (Gaussian linear unit) activation in between. Transformer layers are stacked with residual connection around every two sub-layers. Like BERT4Rec, the output layer $L$ in HydraRec receives the final output $H^L$ for all items of the input sequence. We train the model on a cloze task [13] i.e. in each sequence a randomly masked item must be predicted. To produce a distribution over the target items a 2-layer FFN with GELU activation is used again.

We focus our analysis on encoder-based models, specifically BERT4Rec architecture. The original BERT [5] model was trained for two tasks (next item/token prediction and next sentence prediction), however we narrow our training to the task of next item prediction only similar to [13]. Training bi-directional models for RS tasks can lead to the prediction task becoming trivial as the model will not learn anything useful, therefore we adopt the strategy outlined by [13] where some percentage of the sequence items are masked (just like the masked language model in BERT [5]) and used as labels for the learning task. Therefore, the loss for each masked sequence is the negative log-likelihood of the masked targets. Additionally, we consider two strategies for masking items - unidirectional and bidirectional. Based on this, we refer to the two variants of HydraRec as HydraRecUni and HydraRecBi.

## 4.1 Experiments

In this section we describe the experiments conducted with real-world datasets to establish the overall performance of HydraRec. In summary we answer the following research questions:

*RQ1: How is the overall accuracy of HydraRecUni and HydraRecBi in comparison to scaled dot-product attention in a transformer architecture e.g. BERT4Rec?*

*RQ2: How do other linear attention models compare with HydraRec in accuracy?*

*RQ3: How does the runtime of HydraRec compare to other models?*

*RQ4: How does changing the number of heads affect the accuracy of Hydra Rec?*

Here we define accuracy as the value of the two-evaluation metrics for the validation set. Given that this model builds on a variant of Linear attention, our focus is to compare its effectiveness for the LSR task against prominent linear attention-based models. BERT4Rec (with scaled dot-product attention) is the baseline model. Other models are built on the same architecture with the only update in attention mechanism calculations. We test both variants of HydraRec for accuracy. Theoretical complexity of HydraRec can be linear in embedding and sequence dimensions under certain cases. To analyze this experimentally, we capture the system runtime for training of each of these models.

## 4.2 Datasets

We evaluate the proposed set of models on three real-world representative datasets:

1) ML-1m: MovieLens (1 million) - This dataset has 1 million movie ratings and is a popular choice for recommender system problems.
2) ML - 20m: MovieLens (20 million) - This is the 20 million ratings version of the MovieLens data.
3) Beauty: Amazon Beauty Rating - This is a ratings dataset for beauty related products sold on the Amazon Website.

Movie Lens datasets have longer sequence length for each user. Beauty dataset has shorter sequences in general. A similar pre-processing strategy is used for all three datasets. For each user we create a sequence of items based on the timestamp of the rating (chronological order). Because some of these sequences can be extremely large, we put an upper limit to the sequence length as a hyperparameter. We experimented with different sequence lengths. Padding with zeros was used for sequences shorter than the maximum length.

**Table 1.**     Statistic of the datasets summarized

| Dataset | #Users | #Items | #Interactions | Sparsity |
|---------|--------|--------|---------------|----------|

| ML-1m | 6040 | 3416 | 1 million | 93% |
|-------|------|------|-----------|-----|
| ML-20m | 138,493 | 26,744 | 20 million | 99.4% |
| Beauty | 40,226 | 54,542 | 0.35 million | 99.8% |

### 4.3 Task and Evaluation Metrics

The task at hand is leave-one-out evaluation (next item prediction) like [13][24]. For every user the last item is used for testing. As described in section 3.3 the training is done for a cloze task with some items randomly masked (with an input masking probability of 10 percent for all experiments). Because an LSR task involves the next item prediction, a special token for 'mask' is attached at the end of each sequence that the model predicts. The two variants of HydraRec use two different strategies for masking as described above. We use two well-known evaluation metrics that are used in RS research [30] : (1) Normalized Discounted Cumulative Gain (NDCG) for top k (10), 2) Hit Ratio (HR) for top k (10) recommendations. A higher value indicates a better performance for both these evaluation metrics.

### 4.4 Implementation Details

All parameters are initialized using Gaussian distribution. For HydraRec, we set the hyper-parameters as suggested by [31], including Transformer Layer as 2, attention head equal to 8 and inner FFN layer as 256. We experimented with embedding dimension of 16, 32, 64 and 128 and sequence lengths of 10, 20, 30, 50, 80 and 100. For HydraRec we also experimented with the number of heads equal to the number of embedding dimensions ranging from 16,32,64, 256. The maximum number of training epochs was 200. We report the best results on a validation set (90-10 split) captured for each configuration of the hyperparameters. All experiments are conducted on L4 GPU.

### 4.4 Performance Comparison

In this section we present the results of our experiments. Each experiment was conducted with a batch size varying from 16 to 256, maximum sequence length from 16 to 100 and number of training epochs ranging from 32 to 200. Here the maximum sequence length is a representation of user buying/watching sequence. This has greater relevance for the next item prediction tasks because user buying history impacts what the customer selects next. We are experimenting with cloze tasks. The results for each model under all hyperparameters were recorded and the best values are presented in Table 2 and Table 3. Additionally, we are calling all models using the decomposition technique (equation 5) as linear models (LM) followed by a number to maintain order. HydraRec (equation 6) is a model that is linear in token dimensions and also achieves linearity for some special cases (number of heads becomes equal to embedding dimensions). We experimented with the original Hydra attention model presented in [11] only and incorporated that in the BERT4Rec architecture. Furthermore, the original work describing Hydra attention suggests different kernels to be used in equation (6). We experimented with the original cosine similarity kernel only because in the original paper that achieved highest accuracy.LM1 is the model in [32] in which authors perform decomposition like equation (3) and calculate scaling and matrix multiplication of the query and key-value separately. LM2 refers to the model in [33] that uses linear attention to improve auto-regressive transformers. LM3 [10] is the LinRec model. It is worth noting that the original model uses many other updates to the BERT4Rec architecture other than L2-Norm for attention. In our experiments we did not incorporate any other changes except the attention calculations because the goal is to study the impact of changing attention calculations only. When testing HydraRecUni we applied causal masking to all models including the original dot-product attention. This is different from the original BERT4Rec model.

**Table 2.**      Results on the Validation Set for Causal Masked Models in a
BERT4Rec Architecture

| Dataset | Metric | Dot Product Attention | HydraRecUni | LM1 | LM2 | LM3 | Epochs |
|---------|--------|------------------------|-------------|------|------|------|--------|
| ML-1m | NDCG@10 | 0.4481 | **0.4841** | 0.4751 | 0.4514 | 0.4612 | 180 |
|  |  | 0.4420 | **0.4899** | 0.4755 | 0.4511 | 0.4599 | 200 |
|  | HIT@10 | 0.6901 | **0.7104** | 0.7012 | 0.6802 | 0.7011 | 180 |
|  |  | 0.6992 | **0.7189** | 0.7103 | 0.7024 | 0.7101 | 200 |
| ML-20m | NDCG@10 | 0.6712 | **0.7297** | 0.7193 | 0.6921 | 0.6933 | 50 |
|  |  | 0.6916 | **0.7311** | 0.7274 | 0.7015 | 0.7012 | 100 |
|  | HIT@10 | 0.9012 | **0.9274** | 0.8945 | 0.9012 | 0.9211 | 50 |
|  |  | 0.9101 | **0.9327** | 0.9121 | 0.9145 | 0.9294 | 100 |
| Beauty | NDCG@10 | 0.3921 | **0.4215** | 0.4232 | 0.4115 | 0.4215 | 50 |
|  |  | 0.4212 | **0.4962** | 0.4892 | 0.4731 | 0.4761 | 100 |
|  | HIT@10 | 0.6599 | **0.7012** | 0.6910 | 0.7038 | 0.6961 | 50 |
|  |  | 0.6781 | **0.7135** | 0.7104 | 0.7098 | 0.7001 | 100 |

**Table 3.**      Results on the Validation Set for Bi-directional Models in BERT4Rec
architecture

| Dataset | Metric | Bert4Rec | HydraRecBi | LM1 | LM2 | LM3 | Epochs |
|---------|--------|----------|------------|------|------|------|--------|
|  | NDCG@10 | **0.5087** | 0.4773 | 0.4561 | 0.4567 | 0.4568 | 180 |
| ML-1m | | **0.5270** | 0.4972 | 0.4645 | 0.4651 | 0.4621 | 200 |
| | HIT@10 | **0.7190** | 0.7143 | 0.7103 | 0.6993 | 0.6991 | 180 |
| | | **0.7440** | 0.7202 | 0.7190 | 0.7032 | 0.7011 | 200 |
| ML-20m | NDCG@10 | **0.7561** | 0.7553 | 0.7231 | 0.7434 | 0.7341 | 50 |
| | | **0.7980** | 0.7931 | 0.7525 | 0.7499 | 0.7649 | 100 |
| | HIT@10 | 0.9237 | **0.9239** | 0.8673 | 0.8732 | 0.9012 | 50 |
| | | 0.9491 | **0.9496** | 0.9032 | 0.9101 | 0.9136 | 100 |
| Beauty | NDCG@10 | 0.3012 | 0.4012 | 0.3209 | 0.3087 | 0.3124 | 50 |
| | | 0.3413 | 0.4962 | 0.3312 | 0.3214 | 0.3196 | 100 |
| | HIT@10 | 0.3421 | 0.7612 | 0.6509 | 0.6412 | 0.6318 | 50 |
| | | 0.3498 | 0.7691 | 0.6651 | 0.6417 | 0.6341 | 10 |

**RQ1 and RQ2:** Table 2 and 3 provide the comparison of both models for the two-evaluation metrics i.e. NDGC@10 and Hit@10. The evaluation is done for 3 well-known datasets in recommender systems research. HydraRecUni is compared with dot-product attention and 3 other linear models. Its performance is comparable to other linear models when causal masking is applied. It outperforms the dot-product attention model on all three datasets as measured by the evaluation metrics. The 20M MovieLens dataset and the Amazon Beauty datasets are extremely sparse. HydraRecUni outperforms all other models on these 2 datasets. Under the bi-directional context models, HydraRecBi has a comparable performance with Bert4Rec on NDCG. It does significantly better than other linear models on both evaluation metrics across all three datasets. As the sparsity and size of the dataset begins to increase HydraRecBi gets a higher hit rate than BERT4Rec in some cases. It was also found that this improvement in hit rate increases with embedding dimensions. However, there is a decline in the HydraRecBi hit rate when considering longer sequences. On Amazon beauty dataset, HydraRecUni outperforms all other models. Accuracy improves with increased training time for all datasets and all models. The improvement is incremental in all linear models including HydraRec. For example, for the 1M MovieLens data, NDCG improves by one to two points, but the training time is doubled.

**RQ3:** Of particular interest is the overall runtime of HydraRec when compared to the baseline model for both causal masking scenario versus bi-directional scenario. We study the training time of both variants of the model and compare it against the baselines (dot-product attention models). Figure 2 and Figure 3 show the difference in time taken by HydraRec with the baseline under both scenarios beginning to grow as the number of epochs is increased. These experiments are conducted by considering a fixed sequence length of 20 and embedding dimension of 256. However, a similar trend was recorded for other sequence lengths. The number of epochs impact on the overall accuracy of the model as is evident from the results in Table 2 and 3. This is an important result that shows the effectiveness of HydraRec over the existing dot-product attention-based models under scenarios when high prediction accuracy is required. HydraRec's performance for the casual masked case is better than all other models. For the bi-directional case it shows some loss of accuracy for some scenarios but there is a gain in running time. This can be attributed to the linearity of the attention calculation.

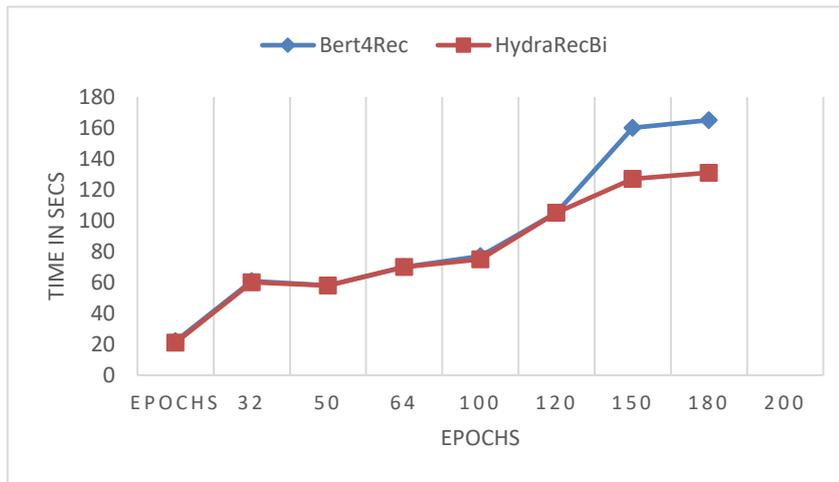

**Fig 2:** Training Time comparison for BERT4Rec Vs HydraRecBi

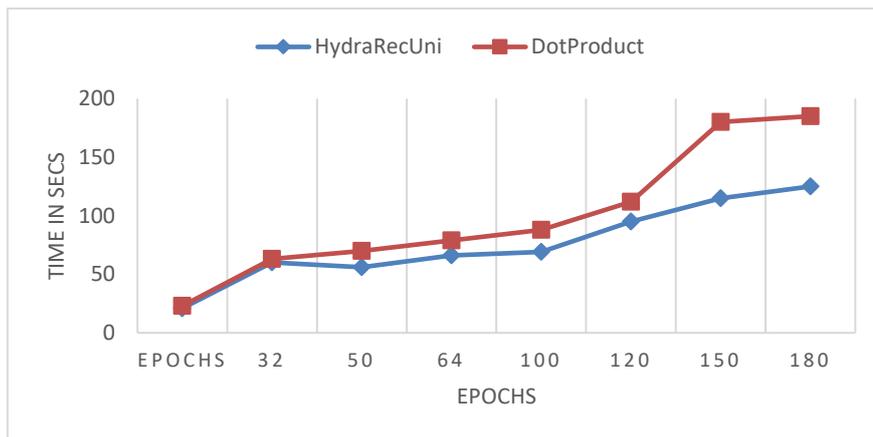

**Fig 3:** Training Time comparison for Dot product Attention Vs HydraRecUni

**RQ4:** In this section we study the impact of changing the number of heads and embedding dimensions on HydraRecBi. In this section we further investigate the impact of changing the number of heads and embedding dimensions on the performance of the HydraRec model. Because the goal is to study the overall trend, the experiments are conducted with only one dataset (ML-1m). We experimented with epochs ranging from 10 to 20 and report the best value here. Batch size is maintained at 32 for all experiments and the maximum sequence length varies from 10 to 64.

**Table 4.**    Impact of number of heads on HydraRec Performance for MovieLens-1m

| Metric | h = 8, d = 64 | h= 8, d = 128 | h= 8, d= 256 | h= 8, d= 512 |
|---|---|---|---|---|
| NDCG@10 | 0.30275166 | 0.3497656 | 0.4113205 | 0.4875293 |
| HIT@10 | 0.54089403 | 0.5975165 | 0.6591059 | 0.6272131 |
| Recall@10 | 0.55129422 | 0.5985125 | 0.6524834 | 0.6554100 |
| **Metric** | **h = 16, d = 64** | **h= 16, d = 128** | **h= 16, d= 256** | **h= 16, d= 512** |
| NDCG@10 | 0.2558698 | 0.2712226 | 0.2831344 | 0.2945021 |
| HIT@10 | 0.4737704 | 0.4868852 | 0.5081967 | 0.5180327 |
| Recall@10 | 0.4918032 | 0.4950163 | 0.5121311 | 0.5192450 |
| **Metric** | **h = 64, d = 64** | **h= 64, d = 128** | **h= 64, d= 256** | **h= 64, d= 512** |
| NDCG@10 | 0.2921790 | 0.2830130 | 0.36593147 | 0.3568240 |
| HIT@10 | 0.4856711 | 0.4943708 | 0.60132450 | 0.5798013 |
| Recall@10 | 0.4895525 | 0.4941388 | 0.61314301 | 0.5698224 |

## 4.5 Discussion

HydraRec is a transformer-based attention model that is linear in sequence and model complexity. This model is used for the next item prediction task of the LSR problem. When comparing the model with existing linear models along accuracy, HydraRecUni outperforms every model. The HydraRecBi variant gives comparable performance to other linear models and sometimes a better performance than dot-product attention. The key advantage of HydraRec lies in complexity improvement and hence a significant reduction in runtime for problems that require longer training times. This is particularly beneficial for problems involving longer sequences, for example when a lot of customer data on an online shopping website (like Amazon) is used to recommend items for next purchase. Additionally, data sparsity is a known issue for recommender system problems leading to lower accuracy and increased runtime. HydraRecUni is a unidirectional model with linear attention. For problems that require the temporal context of a customer buying pattern but can compromise on some contextual information by only taking unidirectional sequence into consideration, this model can be useful. However, for cases where efficiency matters more than accuracy and also for models involving sparse datasets that need to be trained frequently, HydraRec is a clear winner. HydraRec becomes linear in model dimensions when the number of heads equals number the model dimensions. This result can be useful for applications of the LSR problem where efficiency is necessary.

Much of the loss in accuracy of HydraRec can be attributed to the linearity in attention calculations that can lead to loss in context calculations which are more pronounced in longer sequences. Comparing HydraRec with other linear models when used within the same transformer architecture shows a comparable performance. It is worth noticing that HydraRec has a better Hit rate and NDCG for all three datasets than most Linear Models. There was an improvement in performance when the embedding dimensions were increased for all models including HydraRec. HydraRec achieves linearity in complexity when the number of heads equals the number of embedding dimensions. Table 4 shows one such scenario when both are equal to 64. We checked other such cases with h and d equal to 8,16,32 and 64. NDCG@10 was comparable for all cases but Hit and Recall improved with increasing h and d. This is mostly attributed to an increase in embedding dimensions that lead to better contextual representations.

## 5       Conclusion and Future Work

In this study we developed a novel sequential recommendation model called HydraRec, to be used for the LSR problem. The model has computational complexity which is linear in sequence length and for certain cases it can be linear in embedding dimensions. Linearity is achieved via a decomposition technique of the attention model which leads to a model linear in sequence length and further by increasing the number of heads linearity in embedding dimensions can be achieved. We tested the performance of this model by incorporating it in BERT4Rec architecture. Two variants of HydraRec capture two modeling scenarios of the LSR problem. HydraRecUni utilizes the sequence of item selected by a user to predict the next item by using causal masking for future items. HydraRecBi, on the other hand, uses bi-directional context calculations similar to the BERT4Rec model.

HydraRec is theoretically and experimentally more efficient, as is evident from the actual runtime recorded. The accuracy of Hydra attention as compared with other Linear attention mechanisms is better when used for the LSR task. If the task involves causal masking, then HydraRec outperforms dot-product based attention model as well. For a bi-directional model its accuracy is comparable and sometimes better than BERT4Rec. A significant achievement of this model is the saving in runtime which becomes prominent as the number of epochs increases. This points to the cases when model accuracy is important hence training time is more. Under such scenarios HydraRec can be more useful than other linear models. One important constraint of all the models used in this study is that they are used under an encoder transformer architecture. An important addition to this work could be to experiment with decoder architectures as well to better understand the overall effectiveness of Hydra attention in recommender-systems. We experimented with sequence length and recorded results that were the best across all evaluation metrics. A dedicated ablation study for hyperparameters like the impact of sequence length using all datasets from different domains on the performance of HydraRec is an important consideration for future work. Another direction for future work is to consider other item features like category, price etc. when creating item tokens rather than item ids only. Another category of LSR problems involves the next basket prediction. We did not train either of the HydraRec variants for this task. Given the runtime efficiency of HydraRec, this could be a useful application for the model.

# References

[1]     J. Li, P. Ren, Z. Chen, Z. Ren, T. Lian, and J. Ma, "Neural attentive session-based recommendation," in *Proceedings of the 2017 ACM on Conference on Information and Knowledge Management*, 2017, pp. 1419–1428.

[2]     W.-C. Kang and J. McAuley, "Self-attentive sequential recommendation," in *2018 IEEE international conference on data mining (ICDM)*, 2018, pp. 197–206.

[3]     J. Yang, J. Xu, J. Tong, S. Gao, J. Guo, and J. Wen, "Pre-training of Context-aware Item Representation for Next Basket Recommendation."

[4]     A. Vaswani *et al.*, "Attention is all you need," *Adv. Neural Inf. Process. Syst.*, vol. 2017-Decem, no. Nips, pp. 5999–6009, 2017.

[5]     J. Devlin, M. W. Chang, K. Lee, and K. Toutanova, "BERT: Pre-training of deep bidirectional transformers for language understanding," *NAACL HLT 2019 - 2019 Conf. North Am. Chapter Assoc. Comput. Linguist. Hum. Lang. Technol. - Proc. Conf.*, vol. 1, no. Mlm, pp. 4171–4186, 2019.

[6]     B. Zhuang, J. Liu, Z. Pan, H. He, Y. Weng, and C. Shen, "A survey on efficient training of transformers," *arXiv Prepr. arXiv2302.01107*, 2023.

[7]     S. Wang, B. Z. Li, M. Khabsa, H. Fang, and H. Ma, "Linformer: Self-Attention with Linear Complexity," vol. 2048, no. 2019, 2020, [Online]. Available: http://arxiv.org/abs/2006.04768.

[8]     M. Zaheer *et al.*, "Big bird: Transformers for longer sequences," *Adv. Neural Inf. Process. Syst.*, vol. 33, pp. 17283–17297, 2020.

[9]     Z. Ni, H. Yu, S. Liu, J. Li, and W. Lin, "BasisFormer: Attention-based Time Series Forecasting with Learnable and Interpretable Basis," *Adv. Neural Inf. Process. Syst.*, vol. 36, no. NeurIPS, 2023.

[10]    L. Liu *et al.*, "LinRec: Linear Attention Mechanism for Long-term Sequential Recommender Systems," *SIGIR 2023 - Proc. 46th Int. ACM SIGIR Conf. Res. Dev. Inf. Retr.*, pp. 289–299, 2023, doi: 10.1145/3539618.3591717.

[11]    D. Bolya, C. Y. Fu, X. Dai, P. Zhang, and J. Hoffman, "Hydra Attention: Efficient Attention with Many Heads," *Lect. Notes Comput. Sci. (including Subser. Lect. Notes Artif. Intell. Lect. Notes Bioinformatics)*, vol. 13807 LNCS, pp. 35–49, 2023, doi: 10.1007/978-3-031-25082-8_3.

[12]    M. Pagliardini, D. Paliotta, M. Jaggi, and F. Fleuret, "Faster causal attention over large sequences through sparse flash attention," *arXiv Prepr. arXiv2306.01160*, 2023.

[13]    F. Sun *et al.*, "Bert4rec: Sequential recommendation with bidirectional encoder representations from transformer," *Int. Conf. Inf. Knowl. Manag. Proc.*, pp. 1441–1450, 2019, doi: 10.1145/3357384.3357895.

[14]  S. Zhai *et al.*, "An attention free transformer," *arXiv Prepr. arXiv2105.14103*, 2021.

[15]  H. Touvron, M. Cord, M. Douze, F. Massa, A. Sablayrolles, and H. Jégou, "Training data-efficient image transformers & distillation through attention," in *International conference on machine learning*, 2021, pp. 10347–10357.

[16]  G. Shani, D. Heckerman, R. I. Brafman, and C. Boutilier, "An MDP-based recommender system.," *J. Mach. Learn. Res.*, vol. 6, no. 9, 2005.

[17]  F. Mourchid, J. Ben Othman, A. Kobbane, E. Sabir, and M. El Koutbi, "A Markov chain model for integrating context in recommender systems," in *2016 IEEE global communications conference (GLOBECOM)*, 2016, pp. 1–6.

[18]  Y. Yang, H.-J. Jang, and B. Kim, "A hybrid recommender system for sequential recommendation: combining similarity models with markov chains," *IEEE Access*, vol. 8, pp. 190136–190146, 2020.

[19]  A. Graves and A. Graves, "Long short-term memory," *Supervised Seq. Label. with Recurr. neural networks*, pp. 37–45, 2012.

[20]  C.-Y. Wu, A. Ahmed, A. Beutel, A. J. Smola, and H. Jing, "Recurrent recommender networks," in *Proceedings of the tenth ACM international conference on web search and data mining*, 2017, pp. 495–503.

[21]  H. Daneshvar and R. Ravanmehr, "A social hybrid recommendation system using LSTM and CNN," *Concurr. Comput. Pract. Exp.*, vol. 34, no. 18, p. e7015, 2022.

[22]  T. Donkers, B. Loepp, and J. Ziegler, "Sequential user-based recurrent neural network recommendations," in *Proceedings of the eleventh ACM conference on recommender systems*, 2017, pp. 152–160.

[23]  B. Hidasi and A. Karatzoglou, "Recurrent neural networks with top-k gains for session-based recommendations," in *Proceedings of the 27th ACM international conference on information and knowledge management*, 2018, pp. 843–852.

[24]  J. Tang and K. Wang, "Personalized top-n sequential recommendation via convolutional sequence embedding," in *Proceedings of the eleventh ACM international conference on web search and data mining*, 2018, pp. 565–573.

[25]  S. Zhang, L. Yao, A. Sun, and Y. Tay, "Deep learning based recommender system: A survey and new perspectives," *ACM Comput. Surv.*, vol. 52, no. 1, pp. 1–38, 2019.

[26]  D. Bahdanau, K. Cho, and Y. Bengio, "Neural machine translation by jointly learning to align and translate," *arXiv Prepr. arXiv1409.0473*, 2014.

[27]  A. Hendy *et al.*, "How good are gpt models at machine translation? a comprehensive evaluation," *arXiv Prepr. arXiv2302.09210*, 2023.

[28]  L. Wu, S. Li, C.-J. Hsieh, and J. Sharpnack, "SSE-PT: Sequential recommendation via personalized transformer," in *Proceedings of the 14th ACM conference on recommender systems*, 2020, pp. 328–337.

[29]  K. Islam, "Recent advances in vision transformer: A survey and outlook of recent work," *arXiv Prepr. arXiv2203.01536*, 2022.

[30]  G. Adomavicius and A. Tuzhilin, "Toward the next generation of recommender systems: A survey of the state-of-the-art and possible extensions," *IEEE Trans. Knowl. Data Eng.*, vol. 17, no. 6, pp. 734–749, 2005.

[31]  Y. Hou, B. Hu, Z. Zhang, and W. X. Zhao, "Core: simple and effective session-based recommendation within consistent representation space," in *Proceedings of the 45th international ACM SIGIR conference on research and development in information*

*retrieval*, 2022, pp. 1796–1801.

[32]    Z. Shen, M. Zhang, H. Zhao, S. Yi, and H. Li, "Efficient attention: Attention with linear complexities," in *Proceedings of the IEEE/CVF winter conference on applications of computer vision*, 2021, pp. 3531–3539.

[33]    A. Katharopoulos, A. Vyas, N. Pappas, and F. Fleuret, "Transformers are rnns: Fast autoregressive transformers with linear attention," in *International conference on machine learning*, 2020, pp. 5156–5165.